\documentstyle[11pt,amssymb]{article}

\def\be{\begin{equation}}
\def\ee{\end{equation}}

\title{{\hfill\small\tt Theor.\,Math.\,Phys.\,{\bf 6},\,156-164\,(1971)}\\
{\hfill}\\
On the completeness of a system of\\
coherent states}
\author{A.M. Perelomov\\
{\small\em Institute of Theoretical and Experimental Physics,}\\
{\small\em 117 259 Moscow, USSR}\\}

\date{}
\begin{document}
\maketitle{}

\begin{abstract}\noindent
Completeness is proved for some subsystems of a system of coherent states.
The linear dependence of states is investigated for the von Neumann type
subsystems. A detailed study is made of the case when a regular lattice on
the complex $\alpha $ plane with cell area $S=\pi $ corresponds to the states
of the system. It is shown that in this case there exists only one linear
relationship between the coherent states. This relationship is equivalent to
an infinite set of identities. The simplest of these can also be obtained
by means of the transformation formulas for $\theta $ functions.
\end{abstract}

\noindent {\bf 1.} In the study of a number of problems of quantum
mechanics (the theory of measurements [1], quantum optics and
radio electronics [2]-[4], and the decay of a quasi-stationary
state [5]) it has recently been found useful to employ a system of
a so-called coherent states. This is a complete system of states
which has found a wide application recently. It is also known that
this system of states is supercomplete, i.e., there are the
subsystems in this system which are themselves complete. An
interesting example of such a system was pointed out by von
Neumann [1] in the connection with the problem of the most exact
simultaneous measurements of the coordinate and the momentum in
quantum mechanics. However, the proof of the completeness of such
a subsystem apparently was not published.

The aim of the present note is to prove the completeness of some
subsystems of the system of coherent states. In the case of the
von Neumann type subsystems, we have also succeeded in
investigating of the linear dependence of the states. It turns out
that in the von Neumann case there exists one and only one linear
relationship between the states of the subsystems. In the Fock [6]
-- Bargmann [7] representation, the relationship is equivalent to
an infinite set of identities for the functions that are analogous
of the $\theta $ functions. The simplest of these identities can
also be obtained by using of the formulas for the transformation
of the $\theta $ functions. I did not find the corresponding
identities in the mathematical literature in the general case.
Finally, the group aspects of this problem are considered briefly.

\medskip\noindent
{\bf 2.} We shall consider an one-dimensional quantum oscillator\footnote
{\,\,For simplicity, we consider the one-dimensional case. It is readily seen
that all results can be directly generalized to the case of $N$ dimensions.}.
Let as first recall the well-known properties of such a system (see, for
example, [2], [3]). The Hamiltonian of the oscillator has the form\footnote
{\,\,In this paper we use a system of units in which $\hbar =m=\omega =1$
($\hbar $ is the Planck constant, and $m$ and $\omega $ are the mass and
the frequency of the particle vibrations, correspondingly).}
\be H=\frac12\left( p^2+x^2\right) =a^+a+\frac12\,, \ee
where $x$ is the coordinate operators and $p=-\,i\frac{d}{dx}$ is the
momentum operator. The operators
\[ a=\frac1{\sqrt{2}}\,(x+ip)=\frac1{\sqrt{2}}\left( x+\frac{d}{dx}\right),
\qquad a^+=\frac1{\sqrt{2}}\,(x-ip)=\frac1{\sqrt{2}}\left(
x-\frac{d}{dx} \right) \] are known as annihilation and creation
operators, correspondingly.

The self-adjoint operator $H$ is associated with the system of its
eigenvectors $|n\rangle $
\be
H\,|n\rangle =E_n\,|n\rangle ,\qquad E_n=n+\frac12,\qquad n=0,1,2,\ldots .\ee
It is well known that this system is complete and orthogonal.

In many cases, it is more convenient to consider the system of coherent
states $|\alpha \rangle $. This is the system of eigenvectors of the
annihilation operator $a$:
\be a\,|\alpha \rangle =\alpha \,|\alpha \rangle .\ee
Let us mention some of the properties of coherent states (for details, we
refer the reader to [2]-[4]). The spectrum of $a$ fills the entire complex
plane. In other words, the state $|\alpha \rangle $ for any complex $\alpha $
can be normalized, i.e., $\langle \alpha |\alpha \rangle =1$. Expanding such
a state with respect to the states $|n\rangle $, we obtain
\be |\alpha \rangle =\exp\left( -\frac12\,|\alpha |^2\right)
\sum _{n=0}^\infty \frac{\alpha ^n}{\sqrt{n!}}\,|n\rangle .\ee
The following identity also holds:
\be \frac1{\pi }\,\int d^2\alpha \,|\alpha \rangle \langle \alpha |=
\sum _{n=0}^{\infty }|n\rangle \langle n|, \ee
from which it follows that the system of coherent states is complete.

However, the states $|\alpha \rangle $ are not orthogonal to one another.
The scalar product of two such states has the form
\be \langle \alpha |\beta \rangle =\exp\left( -\,\frac12\,\left( |\alpha |^2+
|\beta |^2-2\alpha ^*\beta \right) \right),\qquad |\langle \alpha |\beta
\rangle |^2=\exp\left( -\,|\alpha -\beta |^2\right) . \ee

Equation (5) enables to expand an arbitrary state $|\psi \rangle $
with respect to the states $|\alpha \rangle $:
\be
|\psi \rangle =\frac1{\pi }\,\int d^2\alpha \,\langle \alpha |\psi \rangle
|\alpha \rangle .\ee

Note that if the coherent state $|\beta \rangle $ is taken as $|\psi
\rangle $,then equation (7) defines a linear dependence between the different
coherent states. It follows that the system of coherent states is
supercomplete, i.e., it contains the subsystems which are complete.

Using (4) we obtain the equation for $\langle \alpha |\psi \rangle $ in (7):
\be \langle \alpha |\psi \rangle =\exp\left( -\,\frac12\,|\alpha |^2\right)
\psi (\alpha ^*), \ee
where
\be \psi (\alpha )=\sum _{n=0}^\infty \frac{\alpha ^n}{\sqrt{n!}}\,
\langle n|\psi \rangle . \ee
At the same time, the inequality $|\langle n|\psi \rangle |\leq 1$ means that
the function $\psi (\alpha )$ for the normalization state $|\psi \rangle $
is an entire analytic function of the complex variables $\alpha $.
We also have $|\langle \alpha |\psi \rangle |\leq 1$. Therefore, we obtain
a bound on the growth of $\psi (\alpha )$:
\be |\psi (\alpha )|\leq \exp\left( \frac12|\alpha |^2\right) .\ee
The normalization condition can now be written in the form
\be I=\frac1{\pi }\,\int d^2\alpha \,\exp\left(-\,|\alpha |^2\right )
|\psi (\alpha )|^2 =\langle \psi |\psi \rangle . \ee
The expansion of an arbitrary state $|\psi \rangle $ with respect to
coherent states now takes the form
\be |\psi \rangle =\frac1{\pi }\,\int d^2\alpha \,\exp\left( -\frac12
|\alpha |^2\right ) \psi (\alpha ^*)\,|\alpha \rangle . \ee
Thus, we have established a one-to-one correspondence between the vectors
$|\psi \rangle $ of the Hilbert space and the entire analytic functions
$\psi (\alpha )$ for which the integral $I$ in (11) is finite. This
correspondence is established by equations (9) and (12).

\medskip\noindent
{\bf 3.} Now we can turn to the problem of the completeness of subsystems
of coherent states. Following Bargmann [7], we note that since $\psi
(\alpha ^*)$ is analytic, it is sufficient for its determination
to have values of the function $\psi _i=\psi (\alpha _i^*)$ at points
$\alpha _i^*$ $(i=1,2,3,\ldots )$ whose a sequence has a limit point.
In particular, if $\psi _i=0$, (i.e., $\langle \alpha _i|\psi \rangle =0$),
then $\psi (\alpha ^*)\equiv 0$ and hence $|\psi \rangle =0$. This means that
the system of states $|\alpha _i\rangle $, $i=1,2,3,\ldots $, such that the
sequence $\alpha _i$ has a limit point being complete. Note also that these
arguments go through for any other system $|\beta _k\rangle $ which differs
from the $|\alpha _k\rangle $ system only by the removal of a finite number
of states. It follows that any such system is supercomplete.

On the $\alpha $ plane let us now consider a set of points $\alpha _i$ that
does not have limit points in a finite part of the plane and the
corresponding
set of coherent states $\{|\alpha _i\rangle \}$. We can always construct
an entire function $\psi (\alpha )$ having zeros at the points $\alpha _i^*$.
At the same time, the state $|\psi \rangle $ constructed by means
of equation (12) will be orthogonal to all
states $|\alpha _i\rangle $. If, in addition, $|\psi \rangle $ is a
square-integrable state, i.e., the integral $I$ in (11) converges, the system
$\{|\alpha _i\rangle \}$ is not complete. However, if any state $|\psi
\rangle $ that is orthogonal to all the states $|\alpha _i\rangle $ is not
square-integrable (i.e., for any entire function with zeros at the points
$\alpha _i^*$ the integral $I$ in (11) diverges), the system
$\{|\alpha _i\rangle \}$ is complete.

Thus, the question of the completeness of the system $\{|\alpha
_i\rangle \}$ is reduced to establishing whether the integral $I$
in (11) converges. In turn, this is determined by the asymptotic
behavior of the function $\psi (\alpha )$ as $|\alpha |\to \infty
$. If the order of the growth $\lambda $ of the entire function
$\psi (\alpha )$ is greater than two, the integral diverges; if
$\lambda <2$, the integral converges.

On the other hand, from the theory of entire functions [8], it is well
known that the order of growth $\lambda $ of an entire function $\psi
(\alpha )$ with zeros at the points $\alpha _i$ is not less than the exponent
of convergence $\lambda _1$ of the sequence $\alpha _i$, $\lambda \geq
\lambda _1$, and it is not difficult to construct functions with $\lambda
=\lambda _1$. We recall that the exponent of convergence of a sequence
$\alpha _i$ is a number $\lambda _1$ such that for the arbitrarily small
$\varepsilon >0$ and $\delta >0$, the series $\sum _n 1/|\alpha _n|^
{\lambda _1+\varepsilon }$ converges and the series $\sum _n 1/|\alpha _n|^
{\lambda _1-\delta }$ diverges. At the same time, the series
$\sum _n 1/|\alpha _n|^{\lambda _1}$ may either converge or diverge. It
follows that for $\lambda _1>2$ the system $|\alpha _n\rangle $ is
supercomplete and for $\lambda _1<2$ such a system is not complete [7].

It remains to consider the more complicated case $\lambda _1=2$. If in this
case the series $\sum _n 1/|\alpha _n|^2$ converges, then, as follows from
the theory of entire functions [8], the function $\psi (\alpha )$ defined
by the canonical product has a minimal type. Consequently, the integral I
in (11) converges and the system of states $|\alpha _n\rangle $
is not complete.

However, if the series $\sum _n 1/|\alpha _n|^2$ diverges, then to obtain
the asymptotic behavior of $\psi (\alpha )$ as $|\alpha |\to \infty $,
we must have a more detailed information about the distribution of the
$\alpha _n$ on the $\alpha $ plane. To this end we introduce two quantities
$\Delta $ and $\delta $. The first of them is the upper density of the
distribution of zeros:
\be \Delta =\overline{\lim_{r\to \infty }}\,N(r)/r^2, \ee
where $N(r)$ is the number of points $\alpha _n$ in a circle of radius $r$.
The quantity $\delta $ characterizes the regularity of the distribution of
points $\alpha _n$ and is defined as follows:
\be \delta =\overline{\lim _{r\to \infty }}\,|\delta (r)|,
\qquad \delta (r)=\sum _{|\alpha _n|<r}\frac1{\alpha _n^2}\,.\ee

Using Lindel\"of's theorem [8], we obtain the following result:
if $\Delta =0$ and $\delta $ is finite, the system $\{|\alpha _n\rangle \}$
is incomplete; however, if $\Delta =\infty $ or $\delta =\infty $, the system
$\{|\alpha _n>\}$ is complete.

The case of $\delta \neq \infty $ and $0<\Delta <\infty $ requires the more
detailed investigation. In this case, $\psi (\alpha )$ is a function of the
second growth order ($\lambda =2$) and a finite type $\mu $. If $\mu _1$
is the minimal type of a function having zeros at the points $\alpha _i^*$,
then for $\mu _1<1/2$ the system $\{|\alpha _n\rangle \}$ is incomplete and for
$\mu _1>1/2$ it is complete. The case $\mu _1=1/2$ requires a special
investigation.

\medskip\noindent
{\bf 4.} In the general case, we have failed to find an algorithm for the
determination of $\mu _1$ in the mathematical literature. However, it is
more interesting the case when the points $\alpha _i$ form a regular lattice
with a cell area $S$ ($\alpha _{mn} =m\omega _1+n\omega _2$, where
$\omega _1$ and $\omega _2$ are complex numbers with $\mbox{Im}\,(\omega _2/
\omega _1) \neq 0$, and $m$ and $n$  are integers) can be studied in fair
details. Namely, we shall prove the following result.

{\bf Assertion 1.} {\em The system of coherent states} $\{|\alpha _{mn}
\rangle \}$, $\alpha _{mn}=m\omega _1+n\omega _2$, $\mbox{Im}\,(\omega _2/
\omega _1) \neq 0$, {\em with a cell area} $S$ {\em is not complete if}
$S>\pi $. {\em It is supercomplete and remains supercomplete if a finite
number of states are removed for} $S<\pi $. {\em Finally, if} $S=\pi $
{\em then the system} $\{|\alpha _{mn}\rangle \}$ {\em is complete and
it remains complete if a single state is removed but becomes incomplete if
any two states are removed.}

Note that for the case of a rectangular lattice, the condition of
completeness of the system $\{|\alpha _{mn}\rangle \}$ with $S=\pi
$ made an important part in the von Neumann investigation [1] on
the question of the most exact simultaneous measurement of the
coordinate and momentum. However, evidently the proof of the
completeness of such a system was not, apparently,  published.

Note also that the $\alpha $ plane is the analog of the phase plane for the
classical oscillator, a cell of the phase plane of area $2\pi \hbar $
corresponding to the cell of the $\alpha $ plane of area $\pi $. The physical
meaning of our assertion becomes clear: the system $\{|\alpha _{k}\rangle \}$
is complete if we have  on the average not less than one coherent state in
the cell of the phase plane of area $2\pi \hbar $.

To prove our assertion, we first construct an entire function
$\psi (\alpha )$  that has zeros at the lattice sites $\beta
_{mn}=\alpha _{mn}^*$. Such a function is well known in the theory
of elliptic functions [9], [10] and is usually denoted as $\sigma
(\alpha )$: \be \sigma (\alpha )=\alpha {\prod _{m,n}}^{\prime
}\left( 1-\frac{\alpha } {\beta _{mn}}\right) \exp\left(
\frac{\alpha }{\beta _{mn}}+\frac12\, \frac{\alpha ^2}{\beta
_{mn}^2}\right) .\ee Here the prime indicates that the product in
(15) is taken over all points of the lattice with the exception of
the point $m=0$, $n=0$. To find the asymptotic behavior of $\sigma
(\alpha )$ as $|\alpha |\to \infty $ we use the identity [9] \be
\sigma (\alpha +\beta _{kl})=(-1)^{kl +k+l}\exp(\eta _{kl}(\alpha
+\beta _{kl}/2))\,\sigma (\alpha ), \ee where \be \eta _{kl}=k\eta
_1+l\eta _2,\qquad \eta _1 =\zeta (\omega _1^*/2),\qquad \eta
_2=\zeta  (\omega _2^*/2), \ee and the function $\zeta (\alpha )$
is defined by \be \zeta (\alpha )=\frac{\sigma '(\alpha )}{\sigma
(\alpha )}=\frac1{\alpha } +{\sum _{m,n}}^{\prime }\left(
\frac1{\alpha -\beta _{mn}}+\frac1{\beta _{mn}} +\frac{\alpha
}{\beta _{mn}^2}\right) . \ee Hence, \be \left| \sigma (\alpha
^*+\alpha _{kl}^*)\right| ^2=\exp\left( \eta _{kl} \alpha ^*+\eta
_{kl}^*\alpha +\frac12\,(\eta _{kl}\alpha _{kl}^*+\eta _{kl}^*
\alpha _{kl})\right) |\sigma (\alpha ^*)|^2. \ee Now we turn to a
new function $\rho (\alpha ,\alpha ^*)\geq 0$: \be |\sigma (\alpha
^*)|^2=\exp\left( \nu {\alpha ^*}^2+\nu ^*\alpha ^2+ 2\mu \alpha
\alpha ^*\right) \rho (\alpha ,\alpha ^*), \ee and require that
this function be doubly periodic, i.e., \be \rho (\alpha +\alpha
_{m,n},\,\alpha ^*+\alpha _{mn}^*)=\rho (\alpha , \alpha ^*). \ee
Hence, we obtain the system of equations for $\nu $ and $\mu $:
\be \nu \omega _1^* +\mu \omega _1=\frac12\,\eta _1,\qquad \nu
\omega _2^*+ \mu \omega _2 =\frac12\,\eta _2, \ee \be \nu {\alpha
_{mn}^*}^2+\nu ^*\alpha _{mn}^2+2\,\mu \alpha _{mn}^* \alpha
_{mn}=\frac12\,(\eta _{mn}\alpha _{mn}^*+\eta _{mn}^*\alpha
_{mn}).\ee From (22) we find the coefficients $\nu $ and $\mu $:
\be \nu =\frac{i}{4S}\,(\eta _1\omega _2-\eta _2\omega _1),\qquad
\mu = \frac{\pi }{2S}\,.\ee Here $S=\mbox{Im}\,(\omega _2\omega
_1^*)$ is the cell area.

Note that equation (23) is automatically satisfied. Now it is
obvious that the asymptotic behavior of $|\sigma (\alpha ^*)|^2$
as $|\alpha |\to \infty $ is determined by the factor $\exp \left(
\nu {\alpha ^*}^2+\nu ^* \alpha ^2+2\,\mu \alpha \alpha ^*\right)
$. Thus, the function $\sigma (\alpha ^*)$ has the second order
and the type $\mu +|\nu |$ ($\mu $ and $\nu $ are defined in
(24)). From equation (20) it is obvious however that $\tilde
\sigma (\alpha ^*)=\sigma (\alpha ^*)\exp\left( -\nu \, {\alpha
^*}^2\right) $ is an entire analytic function of the second order
and of the minimal possible type $\mu $ for the given zeros. Then
\be |\tilde \sigma (\alpha ^*)|^2=\rho (\alpha ,\alpha
^*)\exp\left( 2\nu |\alpha |^2\right) . \ee

Using (16) and (24) we obtain a functional equation for $\tilde
\sigma (\alpha )$: \be \tilde \sigma (\alpha +\alpha
_{kl}^*)=(-1)^{kl +k+l}\,\exp\left( \mu |\alpha _{kl}|^2\right)
\,\exp\left( 2\mu \alpha _{kl}\alpha \right) \tilde \sigma (\alpha
),\qquad \mu =\frac{\pi }{2S}\,.\ee Invoking arguments similar to
those given in the book [10], we conclude that any entire function
satisfying this equation is identical with $\tilde \sigma (\alpha
)$ up to a normalization constant.

The integral $I$ in (11) now takes the form
\be I=\int \rho (\alpha ,\alpha ^*)\,\exp\left( \frac{\pi -S}S\,|\alpha |^2
\right) d^2\alpha .\ee

It is obvious that the integral diverges for $S\leq \pi $ and hence the
system $|\alpha _{mn}\rangle $ is complete. Similarly, for $S>\pi $ this
system is not complete. Note that if the state $|\alpha _{m_0n_0}\rangle $
is removed from the system, then we are led to consider the function
$(1/(\alpha ^*-\alpha ^*_{m_0n_0}))\,\tilde \sigma (\alpha ^*)$. It follows
that for $S<\pi $ the system $|\alpha _{mn}\rangle $ remains complete if
any finite number of states is removed.

If one of the states $|\alpha _{m_0n_0}\rangle $ is removed in the case
$S=\pi $, the question of the completeness of the system of remaining states
reduces to an investigation of the integral
\be I_{m_0n_0}=\int \frac{\rho (\alpha ,\alpha ^*)}{|\alpha -\alpha _
{m_0n_0}|^2}\,d^2\alpha , \ee
which, by virtue of the periodicity of $\rho (\alpha ,\alpha ^*)$, reduces to
$I_{00}$:
\be I_{m_0n_0}=I_{00}=\int \frac{d^2\alpha }{|\alpha |^2}\,\rho (\alpha ,
\alpha ^*)\,. \ee
Here $\rho $ is a non-negative double periodic function all of whose zeros
are situated at the points of the lattice $\beta _{mn}=\alpha _{mn}^*=
m\omega _1^*+n\omega _2^*.$

Let us estimate $I_{00}$. Integrating over a region consisting of
small nonintersecting circles of radius $r_0$ with centers $\gamma
_{mn}$ that coincide with the centers of the lattice
parallelograms, we obtain \be I_{00} > {\sum _{m,n}}^\prime
J_{mn},\qquad J_{mn}=\int \rho (\alpha , \alpha
^*)\,\frac{d^2\alpha }{|\alpha |^2},\quad |\alpha -\gamma
_{mn}|<r_0, \ee but
\[ J_{mn} >\frac{\rho _0\pi r_0^2}{(|\gamma _{mn}|+r_0)^2}>\frac{\pi \rho _0
r_0^2}{4\,|\gamma _{mn}|^2}\,, \]
where $\rho _0$ is the minimum of $\rho (\alpha ,\alpha ^*)$ in the region
$|\alpha -\gamma _{11}|\leq r_0$. Thus,
\be I_{00} > \frac{\pi \rho _0r_0^2}{4}\sum _{m,n}\frac1{|\gamma _{mn}|^2},
\quad \mbox{i.e.}\quad I_{m_0n_0}=I_{00}=\infty . \ee
It follows that if a single state is removed from the system $|\alpha _{mn}
\rangle $, the latter remains complete.

If we remove the states $|\alpha _{m_1n_1}\rangle $ and $|\alpha _{m_2n_2}
\rangle $, we are led to consider the integral
\be \tilde I= \int \frac{\rho (\alpha ,\alpha ^*)\,d^2\alpha }{|\alpha -
\alpha _{m_1n_1}|^2|\alpha -\alpha _{m_2n_2}|^2}\,. \ee
Obviously, this integral converges. Hence, the corresponding system now is
incomplete. Thus, Assertion 1 is completely proved.

As a complete and minimal system we may take  the system of all states
$\{|\alpha _{mn}\rangle \}$ with the exception of the state $|0\rangle $.
It is well known [11] that in the case of a complete and minimal system, we
can always construct an associated complete and minimal system such that
these two systems are bi-orthogonal. Suppose that $\{|\tilde \alpha _{mn}
\rangle \}$ is the system associated with $\{|\alpha _{mn}\rangle \}$ and
it is normalized in the usual manner:
\be \langle \alpha _{mn}|\tilde \alpha _{kl}\rangle =\delta _{mk}\delta _{nl}.
\ee

Using (8), we can readily show that the state $|\tilde \alpha _{kl}\rangle $
corresponds to the function
\be \tilde \psi _{kl}(\alpha ^*)=\exp\left( \frac12\,|\alpha _{kl}|^2\right)
\frac{\alpha _{kl}^*}{\tilde \sigma '(\alpha _{kl}^*)}\,\frac{\tilde \sigma
(\alpha ^*)}{\alpha ^*(\alpha ^*-\alpha _{kl}^*)}\,.\ee
The system $|\tilde \alpha _{kl}\rangle $ enables to obtain a formal
expansion for the arbitrary state $|\psi \rangle $ in a series with respect
to the system of states $|\alpha _{mn}\rangle $:
\be |\psi \rangle \sim {\sum }^{\prime } c_{mn}\,|\alpha _{mn}\rangle , \ee
where
\be c_{mn}=\langle \tilde \alpha _{mn}|\psi \rangle . \ee
However, we should remember that we are dealing with a nonorthogonal
system and the completeness of the system does not imply, for example, the
convergence of the series in (35) [11].\,\footnote{\,\,Ya.G. Sinai kindly
drew my attention to this fact.}

Taking $|\psi \rangle $ to be the vacuum state $|0\rangle $, we obtain
\be |0\rangle \sim {\sum }^{\prime } c_{mn}|\alpha _{mn}\rangle , \ee
where
\be c_{mn}=\langle \tilde \alpha _{mn}|0\rangle =-\,\frac{{\tilde \sigma }'^*
(0)}{{\tilde \sigma }'^* (\alpha _{mn}^*)}\,
\exp\left( \frac{|\alpha _{mn}|^2}2\right) =-\,(-1)^{m+n+mn}\,. \ee
The relationship (37) can now be rewritten in the form
\be \sum _{m,n} (-1)^{m+n+mn}\,|\alpha _{mn}\rangle \sim 0. \ee
To investigate the left-hand side of this relationship we use the
Fock [6]--Bargmann [7] representation. We recall that this  representation
is a concrete realization of the Hilbert space: to the state $|\psi \rangle
=\sum c_{n}|n\rangle $ there corresponds an entire analytic function
$\psi (z)=\sum c_n\,(z^n/\sqrt{n!})$; if $|\psi \rangle =|\alpha \rangle $
is a coherent state, then
\be \psi _\alpha (z)=\exp\left(-\frac12\,|\alpha |^2\right) \exp (\alpha z).
\ee
In the Fock--Bargmann representation, the relationship (39) takes the form
\be f(z)\equiv \sum _{m,n} (-1)^{m+n+mn}\exp\left( -\frac12\,|\alpha _{mn}|^2
+\alpha _{mn}z\right) \sim 0, \ee
where $\alpha _{mn}=m\omega _1+n\omega _2$, and the cell area is $S=\pi $.

Let us investigate some of the simplest properties of $f(z)$. Obviously,
the series (41) converge for all values of $z$ and $f(z)$, and therefore it
is an entire function. Further combining in pairs the terms $(m,n)$
and $(-m,-n)$ we see that $f(z)$ is an even function. Let us consider how
$f(z)$ varies as a result of a shift of the argument by $\alpha _{kl}^*$.
Under such a shift each term of the series (41) is multiplied by
\be \exp\left( \alpha _{mn}\alpha _{kl}^*\right) =\exp\left( \frac12\,
(\alpha _{mn}\alpha _{kl}^*+\alpha _{mn}^*\alpha _{kl})\right)
\exp\left( \frac12\,(\alpha _{mn}\alpha _{kl}^*-\alpha _{mn}^*\alpha _{kl})
\right) .\ee
However, if the cell area is $S=\pi $, then
\be \exp\left( \frac12\,(\alpha _{mn}\alpha _{kl}^*-\alpha _{mn}^*
\alpha _{kl})\right) =(-1)^{ml-nk}, \ee
and we obtain
\begin{eqnarray}
f(z+\alpha _{kl}^*)&=&\sum _{m,n}(-1)^{m+n+mn}\,(-1)^{ml-nk}\,\exp\left(
-\frac12\,|\alpha _{m-k,n-l}|^2\right) \nonumber \\
&\times &\exp\left( \frac12\,|\alpha _{kl}|^2+\alpha _{mn}z\right) .
\end{eqnarray}
Substituting $m\to m'+k$ and $n\to n'+l$, we arrive at the identity
\be f(z+\alpha _{kl}^*)=(-1)^{kl+k+l}\exp\left( \frac12\,|\alpha _{kl}|^2
\right) \exp\left( \alpha _{kl}z\right) f(z). \ee
Note that this functional equation is identical with the corresponding
equation (26) for the function $\tilde \sigma (z)$ when $S=\pi $. Since the
solution of this functional equation is unique (see the remark after equation
(26)), we obtain
\be f(z)=C\delta (z),\ee
where $C$ is a constant. However, as we have shown above, $f(z)$ is an even
function whereas $\tilde \sigma (z)$ is an odd function. Therefore,
$C=0$ and consequently $f(z)\equiv 0$.

Thus, we have proved the following assertion.

\noindent{\bf Assertion 2.} {\em The system of analytic functions
} $\exp (\alpha _{mn}z)$ ($\alpha _{mn}=m\omega _1+n\omega _2$,
$S=\pi $) {\em is linearly dependent. The equation relating these
functions has the form} (41).

Expanding $\exp(\alpha _{mn}z)$ in the Taylor series, we arrive at the
identities
\be \sum _{m,n}(-1)^{m+n+mn}\alpha _{mn}^k\,\exp\left( -\frac12\,
|\alpha _{mn}|^2\right) \equiv 0,\qquad k=0,1,2,\ldots ,\ee
Let us consider what these identities yield in the simplest cases. If $k=0$,
then (47) can be written in the form
\be \sum _{m,n} (-1)^{m+n+mn}\exp\left( -\frac{\pi}2\,(am^2+2bmn+cn^2)\right)
\equiv 0, \ee
where $ac-b^2=1$, $a>0$, $c>0$. This identity can be simplified in the case
of a rectangular lattice when $b=0$:
\be \sum _{m,n} (-1)^{m+n+mn}\exp\left( -\frac{\pi }2\left( c^{-1}m^2+cn^2
\right) \right) =0.\ee
Summing over $n$ and then over $m$ in (49), we obtain
\begin{eqnarray}
&&\theta _3\left( 0,\frac{\tau _1}2\right) \theta _3\left( 0,\frac{\tau _2}2
\right) -\theta _3\left( 0,\frac{\tau _1}2\right) \theta _4\left(
0,\frac{\tau _2}2\right) \nonumber \\
&-&\theta _4\left( 0,\frac{\tau _1}2\right) \theta _3\left( 0,
\frac{\tau _2}2\right) -\theta _4\left( 0,\frac{\tau _1}2\right)
\theta _4\left( 0,\frac{\tau _2}2\right) =0, \end{eqnarray}
where
\be \tau _1=ic,\qquad \tau _2={i}c^{-1},\qquad c>0,\ee
and the functions $\theta _3(0,\tau )$ and $\theta _4(0,\tau )$ are defined
by the equations
\be \theta _3(0,\tau )=\sum _{m=-\infty }^{\infty } \exp\left( i\pi \tau
m^2\right) ,\qquad \theta _4(0,\tau) =\sum _{m=-\infty }^\infty (-1)^m
\exp\left( i\pi \tau m^2\right) .\ee
For brevity, we shall write $\theta _i(\tau )$ instead of
$\theta _i(0,\tau )$.

We now use the well-known formulas [9], [10]
\be
\theta _3(\tau )=\sqrt{\frac{i}\tau }\,\theta _3\left(-\,\frac1{\tau }
\right) ,\qquad \theta _4(\tau )=\sqrt{\frac{i}\tau }\,\theta _2\left(
-\,\frac1{\tau }\right) . \ee
Equation (50) then becomes
\be \theta _3(2\tau _1)\,\theta _3(2\tau _2)-\theta _3(2\tau _1)
\theta _2(2\tau _2)-\theta _2(2\tau _1)\theta _3(2\tau _2)-\theta _2(2\tau _1)
\theta _2(2\tau _2) =0, \ee
where $\tau _1$ and $\tau _2$ are defined in (51).

Using also the doubling formulas [9], [10]
\begin{eqnarray}
\theta _2(2\tau ) &=& \frac1{\sqrt{2}}\,\sqrt{\theta _3^2(\tau )-\theta _4^2
(\tau )}, \\
\theta _3(2\tau ) &=& \frac1{\sqrt{2}}\,\sqrt{\theta _3^2(\tau )+\theta _4^2
(\tau )} \end{eqnarray}
and formulas of the type (53)
\begin{eqnarray}
\theta _3(\tau _2) &=& \sqrt{\frac{i}{\tau _2}}\,\theta _3\left(-\,\frac1
{\tau _2}\right) =\sqrt{-i\tau _1}\,\theta _3(\tau _1),\nonumber \\
&&\\
\theta _4(\tau _2) &=& \sqrt{\frac{i}{\tau _2}}\,\theta _2\left(-\,\frac1
{\tau _2}\right) =\sqrt{-i\tau _1}\,\theta _2(\tau _1),\nonumber
\end{eqnarray}
we obtain
\begin{eqnarray}
&&\sqrt{(\theta _3^2+\theta _4^2)(\theta _3^2+\theta _2^2)}-
\sqrt{(\theta _3^2-\theta _4^2)(\theta _3^2-\theta _2^2)}\nonumber \\
&-&\sqrt{(\theta _3^2-\theta _4^2)(\theta _3^2+\theta _2^2)}-
\sqrt{(\theta _3^2+\theta _4^2)(\theta _3^2-\theta _2^2)}=0.\end{eqnarray}
Since [9], [10], equation (58) can be transformed into an identity
\be \theta _3^4=\theta _2^4+\theta _4^4. \ee
Thus, even the proof of the simplest of the identities (47) requires a
detailed knowledge of the relationships between $\theta $ functions.
The proof of the more complicated identities of the type (47) by direct
methods is evidently a nontrivial problem.

In this paper, we have considered the case of a single degree of freedom.
However, it is readily seen that Eq.(41) can be directly generalized to
the case of $N$ degrees of freedom.

\medskip\noindent{\bf 5.} We are intrigued by the appearance of $\theta $
functions in this problem. In this connection, we should like to mention
the Cartier paper [12] which
establishes a certain connection between the Heisenberg commutation
relations for the coordinate and momentum operators and $\theta $ functions.
Cartier shows that $\theta $ functions arise in an investigation of the
representation of the group $G$ associated with the commutation relations if
one assumes that the representation is induced by a certain representation
of a discrete subgroup $\Gamma $ of $G$\,\,\footnote{\,\,A number of general
questions related to discrete subgroups of continuous groups are discussed
in the book by Gel'fand, Graev, and Pyatetskii-Shapiro [13].}.
However, our approach to the question of $\theta $ functions differs from
that of Cartier.

Let us briefly consider the relationship between these two approaches. It is
readily seen that coherent states can be obtained from the ``vacuum state"
$|0\rangle $ by means of the so-called displacement operators [2], [3]
\be |\alpha \rangle =D(\alpha )\,|0\rangle , \ee
where the unitary operator $D(\alpha )$ is defined by
\be D(\alpha )=\exp\left( \alpha \alpha ^+-\alpha ^*\alpha \right) ,\ee
and $a^+$ and $a$ were defined at the beginning of this paper. The law of
multiplication for the operators $D(\alpha )$ has the form
\be D(\alpha _2)\,D(\alpha _1)=\exp\left( i\,\mbox{Im}\,(\alpha _2\alpha _1^*)
\right) D(\alpha _2+\alpha _1).\ee

It follows that the operators $\exp(2\pi it)\,D(\alpha )$ form a group $G$.
An element $g$ of this group is defined by a real number $t$ and a complex
number $\alpha $: $g=(t,\alpha )$. The product of two group elements
$g=g_2g_1$ is given by
\be g=(t,\alpha ),\qquad t=t_2+t_1+\frac1{2\pi }\,\mbox{Im}\,(\alpha _2
\alpha _1^*),\qquad \alpha =\alpha _2+\alpha _1. \ee
Note that the operators $D(\alpha _2)$ and $D(\alpha _1)$ commute if and only
if the area of the parallelogram spanned by the vectors $\alpha _2$ and
$\alpha _1$ is a multiple of $\pi \colon S=\mbox{Im}\,(\alpha _2
\alpha _1^*)=k\pi $, where $k$ is an integer.

Consider the set of operators $D_{np}=D(\alpha _{np})$, where $\alpha _{np}$
is a point of a regular lattice with the cell area $S=\pi $. These operators
form a discrete commutative group with multiplication law
\be D_{np}D_{kl}=(-1)^{B(n,p;k,l)}\,D_{n+k,p+l},\qquad B(n,p;k,l)=nl-pk.
\ee

Let $\Gamma $ be the discrete subgroup of $G$ consisting of
elements of the form $g=(0,\alpha _{np})$ and $g=(1/2, \alpha
_{np})$. The operators $\pm D_{np}$ form a representation of this
discrete subgroup and the set of states $|\alpha _{np}\rangle $
forms a basis of a certain representation of the group $\Gamma $.
Let us try to release the sign factor in (64). To this end, we go
over to the new operators \be D_{np}=(-1)^{F(n,p)}\,\bar D_{np},
\ee and require that \be \bar D_{np}\,\bar D_{kl} =\bar
D_{n+k,p+l}. \ee This gives the equation for $F(n,p)$: \be
F(n+k,p+l)=F(n,p)+F(k,l)+B(n,p;k,l)(\mbox{mod}\,2) \ee ($B$ is
defined in (64)). This equation is identical with equation (71) in
[12] for $m=1$, $N=1$, and, as it is readily verified, has the
solution \be F(k,l)=kl+k+l. \ee We go over to a new system of
states: \be |\bar \alpha _{kl}\rangle =\bar D_{kl}|0\rangle
=(-1)^{kl+k+l}\,|\alpha _{kl}\rangle , \ee in which the action of
the operator $\bar D_{kl}$ has the form \be \bar D_{kl}\,|\bar
\alpha _{np}\rangle =|\bar \alpha _{k+n,l+p}\rangle .\ee From this
equation we can immediately obtain the relationship between the
states of the system $|\bar \alpha _{np}\rangle $. This is a
relationship of the type $\Sigma c_{np}\,|\bar \alpha _{np}\rangle
\sim 0$. As we know, it is unique. Consequently, it can not be
changed when the operator $\bar D_{kl}$ acts upon it. It is
readily seen that the only relationship satisfying this
requirement has the form \be \sum _{m,n}|\bar \alpha _{mn}\rangle
=\sum _{m,n}(-1)^{mn+m+n}|\alpha _ {mn}\rangle \sim 0 \ee which is
identical with (39).

Note also that the uniqueness of the solution of the functional equations
(26) and (45) for $S=\pi $ follows from the irreducibility of the
representation of $G$ induced by the discrete subgroup $\Gamma $ [12], [13].

I am most indebted to M.I. Graev and Ya.G. Sinai for their interest
in the investigation.

\end{document}